\documentclass[doublespacing]{elsart}

\usepackage{graphicx}

\usepackage{amssymb}
\journal{Journal of Magnetic Resonance}
\begin{document}

\begin{frontmatter}

\title{Spin-lattice NMR relaxation by anomalous translational diffusion}

\author[Bio]{A.E. Sitnitsky\corauthref{cor}},
\ead{sitnitsky@mail.knc.ru}
\author[Phys]{G.G. Pimenov},
\ead{gpimenov@ksu.ru}
\author[Bio]{A.V. Anisimov},
\corauth[cor]{Corresponding author.} \ead{anisimov@mail.knc.ru}

\address[Bio]{Institute of Biochemistry and Biophysics, P.O.B. 30, Kazan
420111, Russia}
\address[Phys]{Physics Department of Kazan State University,
Kremlevskaya 18, Kazan 420008, Russia}

\begin{abstract}

A model-free theoretical framework for a phenomenological
description of spin-lattice relaxation by anomalous translational
diffusion in inhomogeneous systems based on the fractional
diffusion equation is developed. The dependence of the
spin-lattice relaxation time on the size of the pores in porous
glass Vycor is experimentally obtained and found to agree well
with our theoretical predictions. We obtain nonmonotonic behavior
of the translational spin-lattice relaxation rate constant (it
passes through a maximum ) with the variation of the parameter
referring to the
extent of inhomogeneity of the system.\\

\end{abstract}

\end{frontmatter}

\section{Introduction}
Anomalous diffusion processes in randomly disordered media are of
considerable interest at present (see \cite{Met00} and refs.
therein). The list of examples includes, glasses and supercooled
liquids \cite{G92}, porous \cite{Kim97}, \cite{Kim02},
\cite{Dul92}, percolative \cite{Kl97}, \cite{Kl99}, \cite{Kl01},
polymeric \cite{Fi96} and diffusive \cite{Kos03} systems, etc. One
of the powerful methods for investigation of such processes is
that of nuclear magnetic resonance (NMR) diffusometry
\cite{Kim97}. The latter was successfully applied to the systems
listed above [3-9]. The theory of NMR diffusometry in disordered
media \cite{Kim02}, \cite{Kl01} is developed within the framework
of a modern approach to anomalous diffusion  \cite{Met00}
invoking the so called fractional calculus \cite{Ol74}.

The aim of the present paper is to report the experimental results
on spin-lattice relaxation time in porous glass Vycor with
molecules of hexane as a diffusion tracer and to compare them with
phenomenological generalization of the theory of spin-lattice
relaxation in homogeneous media for the case of inhomogeneous one.

A modern description of anomalous diffusion processes in
disordered (e.g., porous) media makes use of the so called
fractional diffusion equation (FDE). The idea of this approach
goes back to many pioneers whose achievements are honored in the
review article \cite{Met00}. The reason for introducing this
equation is as follows. As is well known the mean squared
displacement of a free particle in a homogeneous media increases
linearly with time $<x^2(t)>\ \sim C_1 t$ where $C_1$ is the
diffusion coefficient with the dimension $cm^2/s$. This
conventional Einstein relationship of the classical theory
results from the ordinary diffusion equation for the probability
density function to find the particle at position x at time t and
is a direct consequence of the Fick's second law. In the case of
an inhomogeneous media a disorder (like e.g., obstacles caused by
walls of pores in a porous materials) leads to a slower increase
of the mean squared displacement with time
\begin{equation}
\label{eq1}<x^2(t)>\ \sim C_\alpha t^{\alpha}
\end{equation}
where $C_\alpha$ is the diffusion coefficient with the dimension
$cm^2/s^{\alpha}$ and $0<\alpha\leq 1$ is a phenomenological
parameter characterizing the extent of inhomogeneity. Such
behavior is called sub-diffusion and it originates in any fractal
media due to the presence of dead ends on current ways. It should
be stressed that the departure from the classical theory is of
fundamental character and the latter relationship can not be
reduced to the former one by simple scaling of time and diffusion
coefficient. The most straightforward way to obtain
mathematically sub-diffusion is to generalize the ordinary
diffusion equation by replacing the ordinary derivative in time
by the fractional one of the order $\alpha$. The resulting FDE is
a convenient mathematical tool although a phenomenological one
since there is no lucid and commonly accepted physical meaning of
a fractional derivative at present. Some theoretical motivation
for introducing the FDE arises from its intrinsic relationship
with the so called continuous time random walk theory. In fact
the FDE can be derived from a generalized Langevin equation with
the memory kernel accounting for a power-law waiting time
statistics of the trapping events \cite{Met00}.

Much work is being carried out at present to verify
experimentally the predictions of the approach based on the FDE
\cite{Kim02}, \cite{Kl01}, \cite{Kos03}. As is emphasized in the
papers \cite{Kim02}, \cite{Kl01} NMR microscopy enabled the
authors to verify the anomalous solutions of the FDE for the
first time. The results of these work are compatible with other
approaches to NMR relaxation in porous media
\cite{Kor93},\cite{Kor03}. However the fact that the authors of
\cite{Kim02}, \cite{Kl01} resort to a one-dimensional fractional
counterpart of the ordinary diffusion equation  makes their
theoretical calculations essentially of qualitative character and
leaves room  for further advances. The present paper is a
development along this line. We resort to a three-dimensional FDE
and obtain stringent quantitative relationship of the
spin-lattice relaxation rate constant with the parameters
($\alpha$ and $C_{\alpha}$) of the FDE. Our results provide new
and useful analytical tool to interpret the manifestation of
anomalous translational diffusion in NMR relaxation which is
substantiated by the obtained experimental data.

The matter of NMR relaxation at diffusion through a pore is of
particular significance for biological systems. According to a
widely accepted point of view one of the ways of molecule
penetration through a biological membrane is associated with the
short-lived and long-lived pores with a diameter of several
Engstr\"oms  \cite{Sol68}. Such pores are identified with proteins
- aquaporins revealed in cellular membranes of various organisms
\cite{Ju94},  \cite{Ty99}. In that case the water molecules can
move only one by one according to the so called single file
diffusion. Such pores seem to be mostly suitable for providing
selection and thin regulation of the transfer but the question
about their reality and peculiarities of their functioning still
remains a matter of discussion.

At present the dynamics of molecules of liquids in porous systems
attracts considerable attention with an accent on the questions
of multi-dimensionality of the diffusion and the existence of
bound motion \cite{Pa89}, \cite{Cal84}. The diffusion of water in
pores with a diameter comparable with that of water molecules is
studied by molecular dynamics [17-20]. In \cite{Lev73} it is
shown that fluid mechanics (classical hydrodynamics) can be
qualitatively extrapolated for the one by one  motion of the
water molecules in pores with the diameter of 3 Engstr\"oms.
Water in a narrow pore retains the properties of liquid state
with the diffusion coefficient of approximately $70\%$ of that in
the bulk phase. In \cite{Ai85}, \cite{Ai86} it is shown that
water in narrow pores has the properties of a structured liquid
with angular and radial distributions and with the diffusion
coefficient comparable with that of the bulk phase. According to
the data of neutron diffraction spectroscopy the characteristic
time of passing of the water molecule through the membrane is of
order of 100 mks \cite{Fr80} and thus finds itself within the
time scale of the processes amenable to NMR spectroscopy. It is
reasonable to expect that the constraint of the water molecules'
motion and their interaction with the walls of the pores leads to
peculiarities of magnetic relaxation of water. Our final interest
in transport processes in biological membranes motivates the
present attempt to consider magnetic relaxation by translational
diffusion in the pores of much simpler object namely a porous
glass.

Concluding the Introduction we would like to emphasize the
following fact. The FDE is much more adequate than the ordinary
diffusion equation for the description of processes in
inhomogeneous media but one has to pay a hard cost for gained
facilities. Namely the diffusion coefficient in the FDE acquires
a functional dependence on the phenomenological parameter
$\alpha$ referring to the extent of inhomogeneity. This
dependence is not known in general case and at present one can
only guess some trial possibilities for particular systems of
interest. Besides this diffusion coefficient acquires the unusual
dimension $cm^2/s^{\alpha}$ and thus has minimal crossing (only
at $\alpha =1$) with the ordinary diffusion coefficient with the
dimension $cm^2/s$ which can be obtained by direct measurements
employing the existing scheme of Pulsed-Field Gradient NMR.

\section{Results}
\subsection{Theory}
The theory of spin-lattice relaxation by dipole-dipole interaction
initiated by Bloembergen, Purcell and Pound and developed by
Torrey and Abragam is presented in \cite{Ab61}. The rate constant
of spin-lattice relaxation $1/T_{1}$ is decomposed into
translational and rotational parts
\begin{equation}
\label{eq2}1/T_{1}=(1/T_{1})_{trans}+(1/T_{1})_{rot}
\end{equation}
To calculate this value theoretically one needs to know the
spectral density of the correlation function for spherical
harmonics. The relationship of the value of the spectral density
at a Larmor frequency $\omega _{L}$ with the value of the
contribution to the spin-lattice relaxation rate constant due to
translational diffusion is of the form
\begin{equation}
\label{eq3}(1/T_{1})_{trans}=\frac{3\gamma^4\hbar^2I(I+1)}{2}
\{J^{(1)}(\omega _{L})+ J^{(2)}(2\omega _{L})\}
\end{equation}
where $\gamma$ is the gyromagnetic ratio of the nucleus, $I$ is
their spin and $\hbar$ is the Planck constant. The spectral
densities are proportional to the function $J(\omega)$ of the
spectral density: $J^{(1)}(\omega )=\frac{8\pi}{15} J(\omega )$
and $J^{(2)}(\omega )=\frac{32\pi}{15} J(\omega )$. Thus one
actually needs to calculate the only function $J(\omega)$. The
consideration of the spin-lattice relaxation in homogeneous media
due to diffusion motion of the particles presented in \cite{Ab61}
is based on the ordinary diffusion equation for the probability
density.

How is the value of $1/T_{1}$ altered when we pass from a
homogeneous system to a disordered (e.g., porous) one ? One can
expect that $(1/T_{1})_{trans}$ is very sensitive to the extent
of porosity in the whole range while $(1/T_{1})_{rot}$ becomes to
feel the porosity only when the size of the pores becomes
commensurate with that of the molecules, i.e. in a wide range of
porosity it remains constant. That is why namely the dependence
of the value $(1/T_{1})_{trans}$ on the extent of porosity is of
interest for theoretical consideration in the present paper. We
develop the latter within the framework of a modern approach to
anomalous translational diffusion in disordered systems based on
the FDE. The reason to resort to such generalization of the
ordinary diffusion equation for the description of anomalous
diffusion processes is stated in the Introduction.

The generalization of the ordinary diffusion equation within the
fractional calculus was suggested by Schneider and Wyss
\cite{Sc89}
\begin{equation}
\label{eq4} \frac{\partial P(\bar {r},t)}{\partial t} = C_\alpha
(D_{0 + }^{1 - \alpha } \nabla^2 P)(\bar {r},t)
\end{equation}
\noindent where $P(\bar {r},t)$ is the probability density
function to find the particle at position $\bar {r}$ at time t,
$\nabla^2$ is the three-dimensional Laplace operator, C$_{\alpha
}$ denotes the fractional diffusion constant with the dimension
[cm$^{2}$/s$^{\alpha }$] and  D$^{1-\alpha }_{0 + }$ is the
Riemann-Liouville fractional derivative of order $1-\alpha $ and
with lower limit $0+$ which is defined via the following
relationship \cite{Ol74}
\begin{equation}
\label{eq5} \left( {D_{0 + }^{1-\alpha} f} \right)(x) =
\frac{1}{\Gamma (\alpha )}\frac{d}{dx}\int\limits_0^x {(x -
y)^{\alpha - 1}f(y)dy}
\end{equation}
\noindent where $\Gamma(x)$ is a gamma function. The solution of
equation (\ref{eq4}) for the case of sub-diffusion $0 < \alpha
\le $ 1 with the initial condition $P(\bar {r},0) = \delta (\bar
{r})$ where $\delta (x)$ is a Dirac function is obtained in
\cite{Sc89} and expressed via the Fox's function
\cite{Sr82},\cite{Pr86},\cite{Gl93}
\begin{equation}
\label{eq6} P(\bar {r},t) = \frac{1}{(r^2\pi )^{3 / 2}}H_{12}^{20}
\left( {\frac{r^2}{4C_\alpha t^\alpha }\left| {\begin{array}{l}
 (1,\alpha ) \\
 (3 / 2,1),(1,1) \\
 \end{array}} \right.} \right)
\end{equation}
\noindent The latter is defined as $H_{pq}^{mn} \left( {z\left|
{\begin{array}{l}
 (a_1 ,A_1 ),...,(a_p ,A_p ) \\
 (b_1 ,B_1 ),...,(b_p ,B_p ) \\
 \end{array}} \right.}
 \right)=\frac{1}{2\pi i}\int_Ldsz^{-s}\eta(s)$
where
$\eta(s)=\frac{\prod_{i=1}^m\Gamma(b_i+B_is)\prod_{i=1}^n\Gamma(1-a_i-A_is)}
{\prod_{i=n+1}^p\Gamma(a_i+A_is)\prod_{i=m+1}^q\Gamma(1-b_i-B_is)}$.
The nomenclature in the Fox's function associated with the
vertical bar is explained via its explicit definition by the
contour integral. The requirements to the contour path $L$ are
formulated in \cite{Sr82}.

Now we follow the algorithm of \cite{Ab61} (in what follows all
corresponding results from \cite{Ab61} are obtained as a
particular case $\alpha $=1 of the present approach). Under $\bar
{r}$ we denote the vector $\bar {r}_1 - \bar {r}_2 $ connecting
two identical molecules diffusing relative to each other rather
than the radius-vector of the molecule diffusing relative to a
fixed point. This leads only to the change of $4C_\alpha t^\alpha
$ by $8C_\alpha t^\alpha $ in (\ref{eq6}). Our aim is to calculate
the correlation function
\begin{equation}
\label{eq7} G(t) = N\int\!\!\!\int {\frac{\Upsilon _2^{m^ \ast}
(\theta(0),\varphi(0))}{r_0^3 }} \frac{\Upsilon _2^m
(\theta(t),\varphi(t))}{r^3}P(\bar {r} - \bar {r}_0 ,t)d^3r_0 d^3r
\end{equation}
\noindent where $N$ is the number of spins in 1 cm$^{3}$,
$\Upsilon _n^m(\theta,\varphi)$ is a spherical harmonic and $\ast$
denotes complex conjugate. To be more precise we need the
spectral density of the correlation function $G(t)$ to calculate
the spin-lattice relaxation rate constant with the help of
(\ref{eq2}). At integration in (\ref{eq7}) one should take into
account that $r$ and $r_{0}$ can not be less than some limit
value $d$ -- the least distance to which the molecules can
approach to each other. If the molecules are considered as
spheres of the radius $a$ then $d=2a$ \cite{Ab61}.

Making the Fourier transforming of the function $P(\bar {r} - \bar
{r}_0 ,t)$ with the mentioned above change of $4C_\alpha t^\alpha
$ by $8C_\alpha t^\alpha $ and literally repeating the
manipulation of \cite{Ab61} with spherical functions one obtains
\[
G(t) = \frac{4N}{\pi ^{1 / 2}d^2}\int\limits_0^\infty
{\frac{du}{u^2}} \left[ {J_{\frac{3}{2}} (u)}
\right]^2\int\limits_0^\infty {\frac{dR}{R^2}} \sin \left(
{\frac{uR}{d}} \right)\times
\]
\begin{equation}
\label{eq8} H_{12}^{20} \left( {\frac{R^2}{8C_\alpha t^\alpha
}\left| {\begin{array}{l}
 (1,\alpha ) \\
 (3 / 2,1),(1,1) \\
 \end{array}} \right.} \right)
\end{equation}
\noindent where $J_\nu(x)$ is a Bessel function of order $\nu$ and
$R = \left| {\bar {r} - \bar {r}_0 } \right|$. The integration
can be fulfilled and yields the expressions for both the
correlation function and its spectral density via Fox's functions.
However a Fox's function is to regret not tabulated at present
either in {\it Mathematica} or {\it Maple} or {\it Matlab}. Thus
it is rather difficult to use such formulas for plotting the
behavior of the correlation function and the spectral density.
That is why it is useful to obtain another representation of the
spectral density which enables one to plot the frequency
dependence of this function. For this purpose we make the Fourier
transforming of the Fox function in (\ref{eq8}). The latter can
be achieved with the help of the following trick going back to
original investigations of Fox. In \cite{Sc89} the Mellin
transform of the Fox function is given
\[
M\left\{ {H_{12}^{20} \left( {\frac{r^2}{8C_\alpha t^\alpha }\left|
{\begin{array}{l}
 (1,\alpha ) \\
 (3 / 2,1),(1,1) \\
 \end{array}} \right.} \right)} \right\}(s) =
\]
\begin{equation}
\label{eq9} \frac{1}{\alpha }\left( {\frac{r}{(8C_\alpha )^{1 /
2}}} \right)^{\frac{2s}{\alpha }}\Gamma \left( {\frac{3}{2} -
\frac{s}{\alpha }} \right)\frac{\Gamma (1 - s / \alpha )}{\Gamma
(1 - s)}
\end{equation}
Now we make use of the identity \cite{Sc95}
\begin{equation}
\label{eq10} F_C \left\{ {f(x);\omega } \right\} = M^{ - 1}\left\{
{\Gamma (s)\cos \frac{\pi s}{2}M\left\{ {f(x);1 - s} \right\}}
\right\}
\end{equation}
\noindent where $F_{C}$ denotes the cosine Fourier transform and
$M$ denotes the Mellin transform which for our case is given by
formula (\ref{eq9}). \noindent We denote the characteristic time
\begin{equation}
\label{eq11} \tau _\alpha = \left( {\frac{d^2}{2C_\alpha }}
\right)^{1 / \alpha }
\end{equation}
\noindent As a result we obtain
\[
J(\omega ) = \frac{N\tau _\alpha ^{1 - \alpha } }{C_\alpha
d\left( {\omega \tau _\alpha } \right)^{1 - \alpha }}\sin
\frac{\pi \alpha }{2}\int\limits_0^\infty {du} \left[
{J_{\frac{3}{2}} (u)} \right]^2\times
\]
\begin{equation}
\label{eq12} \frac{u}{u^4 + \left(\omega \tau
_\alpha\right)^{2\alpha}+2u^2\left(\omega \tau
_\alpha\right)^{\alpha}\cos \frac{\pi \alpha }{2}}
\end{equation}
The integral in (\ref{eq12}) can be easily calculated numerically.
We use {\it Mathematica} for this purpose. Formula (\ref{eq12})
is the main theoretical result of the present paper. This formula
is a direct generalization of the textbook result (see e.g.,
\cite{Ab61} formula VIII.114) for the homogeneous case $\alpha=1$
for the case of arbitrary $ 0<\alpha\le1 $. The spectral density
is finite at $\omega \tau _\alpha \to 0$ only for $\alpha $=1. At
all $\alpha < 1 $ it is power-law divergent in this limit.
\subsection{Phenomenology}
The predictions of the present theory are defined by the
dependence of the diffusion coefficient $C_\alpha$ on the
phenomenological model parameter $\alpha $ referring to the extent
of inhomogeneity. This dependence is of most interest because it
characterizes the disordered media \cite{Met00}, \cite{Kl01},
\cite{Kos03}. Its derivation is a separate and still open problem.
It should be a matter of a fundamental theory and is out of the
scope of the present paper. In our phenomenological approach we
restrict ourselves by a model example. As such we choose a
particular behavior of the diffusion coefficient
\begin{equation}
\label{eq13} C_\alpha =\frac{d^2}{2\Gamma(\alpha)}\left(\frac{2C_1
}{d^2}\right)^\alpha
\end{equation}
where $C_1$ is given by the Stokes formula
\begin{equation}
\label{eq14} C_1 =\frac{k_B T}{6\pi a\eta}
\end{equation}
Thus the characteristic time is parameterized as
\begin{equation}
\label{eq15} \tau _\alpha = \tau_1 \Gamma(\alpha)^{1 / \alpha }
\end{equation}
where
\begin{equation}
\label{eq16} \tau _1 = \frac{12\pi a^3\eta}{k_B T }
\end{equation}

The behavior of the spectral density at several values of the
parameter $\alpha $ is depicted in Fig.1. Substitution of
(\ref{eq12}),(\ref{eq13}) and (\ref{eq15}) into (\ref{eq3}) yields
the normalized spin-lattice relaxation rate constant
\[
\frac{5dC_1}{4\pi
N\gamma^4\hbar^2I(I+1)}(1/T_{1})_{trans}(\omega_L) =
\]
\[
\left( {\omega_L\tau _1}\right)^{\alpha-1}\Gamma(\alpha)\sin
\frac{\pi \alpha }{2}\int\limits_0^\infty {d u} \left[
{J_{\frac{3}{2}} (u)} \right]^2u \times
\]
\[
\Biggl\lbrace \frac{1}{u^4+\left(\omega_L
\tau_1\right)^{2\alpha}{\Gamma(\alpha)}^2+2u^2\left(\omega_L\tau
_1\right)^{\alpha}\Gamma(\alpha)\cos\frac{\pi\alpha}{2}} +
\]
\begin{equation}
\label{eq17} \frac{2^{\alpha +1}}{u^4+\left(2\omega_L
\tau_1\right)^{2\alpha}{\Gamma(\alpha)}^2+2u^2\left(2\omega_L\tau
_1\right)^{\alpha}\Gamma(\alpha)\cos\frac{\pi\alpha}{2}}\Biggr\rbrace
\end{equation}

Its behavior on the parameter of inhomogeneity $\alpha $ at
different values of $\omega_L\tau_1$ is depicted in Fig.2. At
experimentally used Larmor frequency $w_L\tau_1\approx0.015$ the
rate constant of spin-lattice relaxation has a maximum at
$\alpha\approx0.35$ (with $w_L\tau_{\alpha}\approx0.21$).

\subsection{Objects of investigation, experimental procedure and results}
Porous glasses of the class ''Vycor'', with characteristics
presented in Table 1, were used as model samples of porous media.
There R is the size of glass microparticles, p is the pore
radius, $S_{n}$ and $V_{n}$ are the pore specific surface area and
volume, respectively. The primary porosity of these glasses is
determined by the pore size p, and the secondary one is determined
by the glass microparticle size R.

In the experiment, hexane vapor was absorbed at the temperature of
$21^{o}$C in the $\Pi$-shaped cell consisting of two connected
tubes with the inner diameter of 7 mm. One tube was filled with a
sample of porous glass to the height of 12 mm, and into the other
tube the liquid hexane was poured. Hexane vapor absorbtion took
place in the presence of a certain amount of air. Prior to cell
soldering, the leg with porous glass was heated at the temperature
of $140^{o}$C to remove water vapor. The amount of the absorbed
hexane was determined from the amplitude of free induction decay
after the $90^{o}$ radio frequency pulse. The calibrating of the
signal was carried out using the reference sample with the known
contents of protons. The process of absorption consists of two
stages. The first faster stage has the character of mono
molecular absorption and is terminated at the amount of absorbed
molecules equal or close to the amount necessary to form a mono
molecular layer. The second slow stage is associated with the
capillary condensation, and lasts from 24 hours (''Vycor 20'') to
a month and a half (''Vycor 200''), and is accomplished by
filling the micropores. A good coincidence of calculated and
measured values of the amount of hexane, necessary to fill
completely the primary pores, is observed. Spin-lattice
relaxation times, $T_{1}$, of absorbed hexane were measured using
a zero-method of the pulse sequence $180^{o}-90^{o}$ on the
coherent NMR relaxation meter using protons at the frequency of
19.5 MHz. Taking the average size of a hexane molecule as $a=5 A$
($d=2a=10 A$) and its diffusion coefficient in bulk water as
$4.17\ast10^{-5} cm^2/s$ at room temperature we obtain
$\tau_1\approx10^{-10}s$ and consequently
$\omega_L\tau_1\approx0.015$.

The decay of the longitudinal magnetization of the hexane in the
region of full filling of the pores is one-exponential for all
porous glasses. As an example the decay of the longitudinal
magnetization of the hexane in the ''Vycor 160'' is presented in
Fig.3. The experimentally observed spin-lattice relaxation times
depending on the inverse radius of the pores are plotted in
Fig.4. One can see that with the variation of the pore size from
400 A to 55 A the rate of longitudinal relaxation increases and
for p=20 A it begins to decrease in accordance with the
prediction of the theory.

\section{ Discussion}
Our theoretical approach is based on the FDE like that of the
papers \cite{Kim02}, \cite{Kl01}. However in these papers a
one-dimensional fractional counterpart of the ordinary diffusion
equation is employed to gain qualitative insight. In contrast we
consider a three-dimensional one that enables us to develop
quantitative formalism. In our approach we do not a priory assume
any type and character of the motion like e.g.,
quasi-two-dimensional bulk mediated surface diffusion of the
absorbed molecules in the pores \cite{Kim02}. Our approach due to
its phenomenological model-free character is flexible enough to
include formally such type of motion as a particular case.

Translational diffusion was previously taken into account  within
the so called reorientation mediated by translational
displacements model \cite{Kim97}, \cite{Kim02}. It is assumed
there that dipolar coupling responsible for the relaxation
mechanism is predominantly of an intramolecular nature, i.e. the
fluctuations causing relaxation are exclusively due to molecular
reorientations \cite{Kim97}. In our approach we emphasize another
facet of translational diffusion manifesting itself in the
intermolecular interactions. We conceive the process as a bulk
three-dimensional diffusion in space with obstacles imposed by
the porous media. In our opinion the porosity of any extent
considerably manifests itself in the translational diffusion. On
the other hand it can affect rotational diffusion appreciably
only when the size of pores becomes commensurate with that of the
molecules. That is why the value of $(T_1)_{rot}$ is independent
on the size of the pores in a wide range of the extent of
porosity while $(T_1)_{trans}$ is very sensitive to this
parameter. The latter value characterizes the porosity of the
system and is of main interest in the present paper.

The most important feature of the spectral density (\ref{eq12})
which distinguishes it in the general case from the particular one
$\alpha =1$ considered in \cite{Ab61} is the divergence of the
spectral density at low frequencies (see Fig.1). Our
phenomenological model-free approach can be compared with that
based on a molecular model of the porous system \cite{Kor93}.
Fig.1 is qualitatively similar to Fig.2 from \cite{Kor93}.
However there are significant quantitative distinctions between
two approaches. The authors of \cite{Kor93} obtain a logarithmic
divergence of the spectral density at low frequencies while we
obtain a power-law one. Our result agrees with the conclusions of
the papers \cite{Kim02}, \cite{Kor03} where a power-law divergence
$1/T_1\propto1/\omega^{\beta}$ at low frequencies is reported.
However the authors of these papers conclude that the proton
relaxation process is mainly due to intramolecular dipolar
interaction while we explore the range of inhomogeneity
(particularly porosity) of the system where in our opinion
intermolecular interactions of molecules determine the dependence
of the relaxation rate constant on translational diffusion
process.

It is known  that a confinement enhances significantly both
spin-lattice and spin-spin relaxation rate constants and alters
their frequency and temperature dependencies (see \cite{Kor93}
and refs. to previous works of its authors). In these papers a
monotonic behavior of the spin-lattice relaxation rate constant
($1/T_1\propto1/p$ or $1/T_1\propto1/p^2$ where p is the average
pore size) is reported. In contrast we obtain nonmonotonic
behavior of the translational spin-lattice relaxation time with
the variation of the parameter $\alpha$ referring to the extent of
inhomogeneity ( or that of porosity, namely the size of the
pores) with $\alpha=1$ corresponding to the case of homogeneous
system. We find that at a given temperature there is an optimal
porosity for maximal spin-lattice relaxation rate constant. Such
behavior is determined by contribution from translational
diffusion. For experimentally used Larmor frequency
$w_L\tau_1\approx0.01$ and for a particular model dependence of
the fractional diffusion coefficient on the parameter of
inhomogeneity  the rate constant of relaxation first increases
with the decrease of $\alpha$ then passes through a maximum at
$w_L\tau_{\alpha}\approx0.1$ and finally sharply decreases. The
reason for such behavior is quite similar to that for ordinary
diffusion in a homogeneous system with the variation of
temperature. In the latter case the characteristic time
$\tau_1(T)$ is a function of temperature and the rate constant of
spin-lattice relaxation has a maximum at some temperature where
$\omega_L\tau_1\approx1$. For inhomogeneous systems the
characteristic time $\tau_\alpha(T,\alpha)$ besides temperature
is a function of the parameter of inhomogeneity $\alpha$ and the
variation of the latter (e.g., that of the porosity of the
system) at constant temperature leads to a maximum of the
spin-lattice relaxation rate constant at some $\alpha$. However
this maximum in general case exists at $\omega_L\tau_\alpha$
different from $1$.

\section{Conclusions}
A generalization of the ordinary diffusion equation within the
framework of the fractional calculus provides natural
phenomenological generalization of the theory of spin-lattice NMR
relaxation in homogeneous systems for the case of the
inhomogeneous ones. This development is carried out within a
general modern trend to extend the theory  of Gaussian
translational diffusion in homogeneous systems for the case of
anomalous diffusion in disordered (e.g., porous) systems by
making use of the fractional diffusion equation. The
translational contribution into spin-lattice relaxation time is
found to be highly sensitive to the extent of inhomogeneity. It
exhibits nonmonotonic behavior with the variation of the
parameter $\alpha$ referring to the extent of inhomogeneity (with
$\alpha=1$ corresponding to the case of a homogeneous and
isotropic system and $\alpha=0$ corresponding to that of an
absolutely inhomogeneous one). The rate constant of relaxation
passes through a maximum with the decrease of $\alpha$. For
experimentally used Larmor frequency $w_L\tau_1\approx0.015$ and
for a particular model dependence of the fractional diffusion
coefficient on the parameter of inhomogeneity the maximum is at
$\alpha\approx0.35$ ($w_L\tau_{\alpha}\approx0.21$). One can
conclude that the present work provides a reliable theoretical
framework for the analysis of anomalous translational diffusion
processes by NMR microscopy.

We obtain the explicit dependence of the rate constant of
spin-lattice relaxation by anomalous translational diffusion on
the parameter of inhomogeneity $\alpha$ taking place in the
fractional diffusion equation. Formula (\ref{eq12}) is a direct
generalization of the textbook result (see e.g. formula VIII.114
in \cite{Ab61}) for the homogeneous case $\alpha=1$ for the case
of arbitrary $ 0<\alpha\le1 $. Our calculations predict that
there occur a maximum of the relaxation rate constant at varying
this parameter. We carry out experimental measurements of the
relaxation rate constant in porous glass
Vycor and obtain data which agree well with our theoretical findings.\\

Acknowledgements. The authors are grateful to Prof. V.D. Fedotov
and R.Kh. Kurbanov for helpful discussions. The work was supported
by the grant from RFBR.

\newpage
\begin{table}[htbp]
\caption{Characteristics of porous glasses.}
\begin{tabular}
{|p{65pt}|p{65pt}|p{65pt}|p{65pt}|p{65pt}|} \hline Designations&
$R$, mm& $p$,\ \ \ $\AA\ \ $$\pm10\%$& $S_{n}$, $m^{2}/g$
$\pm5\%$&
$V_{n}$, $cm^{2}/g$ $\pm5\%$\\
\hline Vycor 20& 0.1 -- 0.3& 20& 200&
0.2 \\
\hline Vycor 55&$<$0.05& 55& 254&
0.7 \\
\hline Vycor 160& 0.1 -- 0.2& 160& 137.5&
1.1 \\
\hline Vycor 220& 0.1 -- 0.3& 220& 118&
1.3 \\
\hline Vycor 400&$<$0.1& 400& 25&
0.5 \\
\hline
\end{tabular}
\label{tab}
\end{table}

\newpage
\begin{figure}
\begin{center}
\includegraphics* [width=\textwidth]{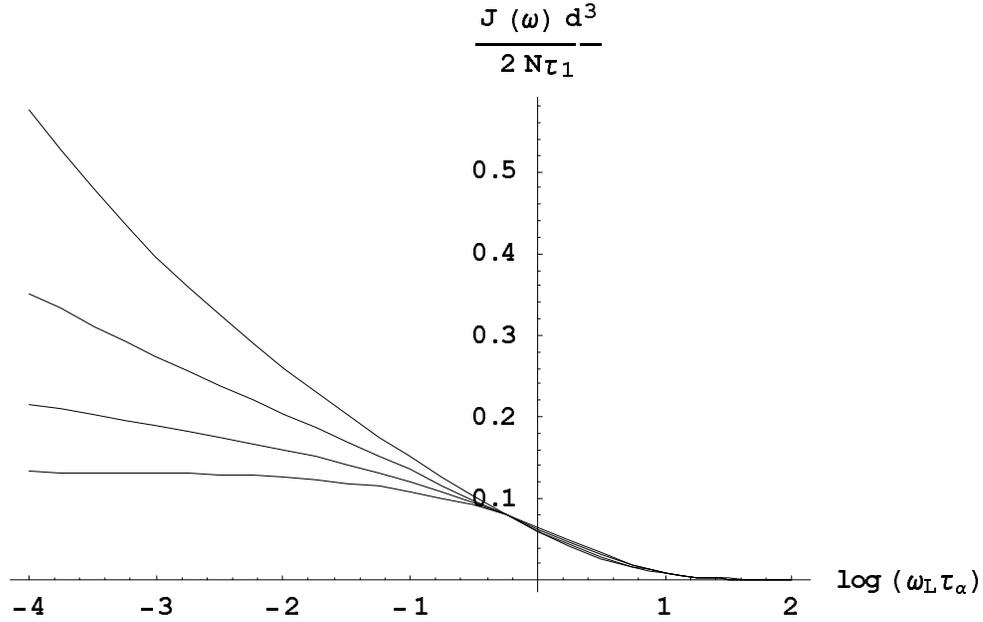}
\end{center}
\caption{The dependence of the spectral density of the
correlation function for spherical harmonics due to anomalous
translational diffusion given by the (\ref{eq12}) on the reduced
Larmor frequency $\omega_L\tau_{\alpha}$ at different values of
the parameter of inhomogeneity $\alpha$. The values of the latter
parameter from the down line to the upper one respectively are:
1; 0.95; 0.9; 0.85. Here $N$ is the number of spins in 1 cm$^{3}$,
$d$ is the doubled radius of the molecules, $\tau_1$ is the
characteristic time for the homogeneous media given by the
(\ref{eq16}) and $\tau_{\alpha}$ is the characteristic time for
the inhomogeneous media given by the (\ref{eq15}).} \label{Fig.1}
\end{figure}

\clearpage
\begin{figure}
\begin{center}
\includegraphics* [width=\textwidth] {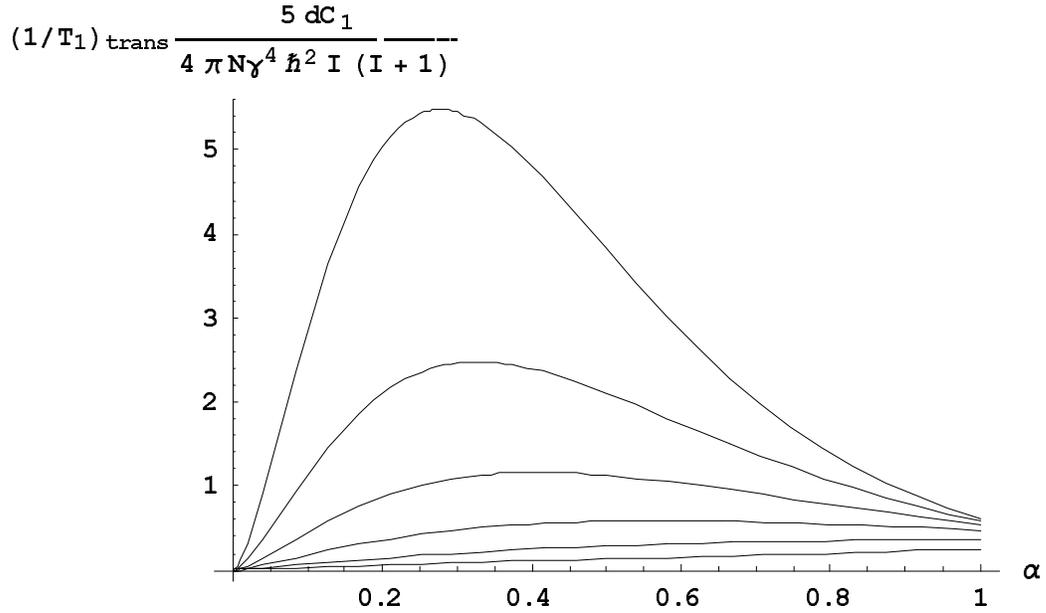}
\end{center}
\caption{The dependence of the contribution to the spin-lattice
relaxation rate constant due to anomalous translational diffusion
given by the (\ref{eq17}) on the parameter of inhomogeneity
$\alpha$ at different values of the reduced Larmor frequency
$\omega_L\tau_1$. The values of the latter parameter from the
down line to the upper one respectively are: 1.; 0.4; 0.16; 0.06;
0.025; 0.01. Here $N$ is the number of spins in 1 cm$^{3}$, $d$ is
the doubled radius of the molecules, $C_1$ is the diffusion
coefficient for the homogeneous media given by the Stokes formula
(\ref{eq14}), $\tau_1$ is the characteristic time for the
homogeneous media given by the (\ref{eq16}), $\gamma$ is the
gyromagnetic ratio of the nucleus, $I$ is their spin and $\hbar$
is the Planck constant.} \label{Fig.2}
\end{figure}

\clearpage
\begin{figure}
\begin{center}
\includegraphics* [width=\textwidth] {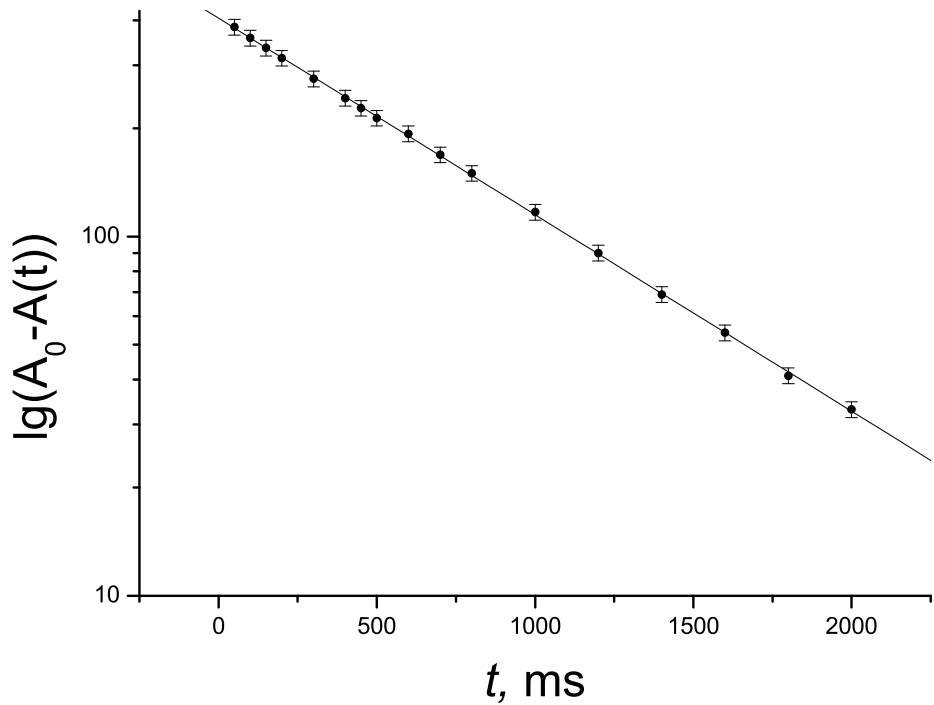}
\end{center}
\caption{The experimental decay of the longitudinal magnetization
for hexane in the porous glass Vycor 160 with the radius of the
pores 160 A. Solid line is the result of the linear fit of the
data. Here t is the current time of the magnetization's
relaxation, $A_0$ is the equilibrium value of the magnetization
and $A(t)$ is the current value of the magnetization.}
\label{Fig.3}
\end{figure}

\clearpage
\begin{figure}
\begin{center}
\includegraphics* [width=\textwidth] {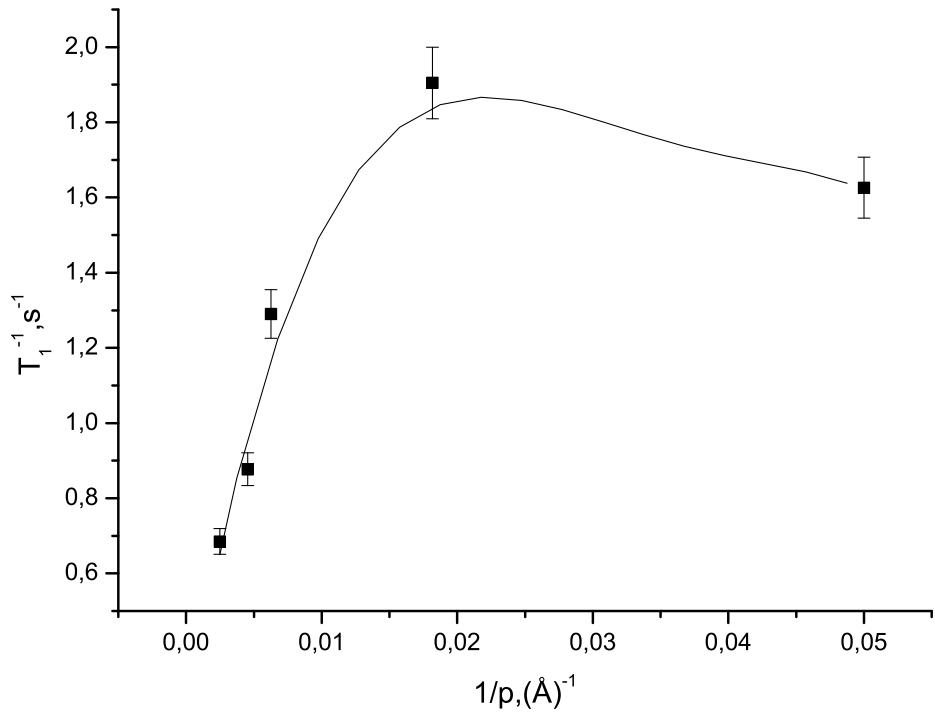}
\end{center}
\caption{The experimental dependence of the spin-lattice
relaxation rate constant on the reciprocal radius of the pores
for the porous glasses Vycor at the Larmor frequency 19.5 MHz.
Here $T_1$ is the spin-lattice relaxation time and p is the
radius of the pores measured in Engstr\"oms \AA. The solid line is
an empirical fitted one "to guide the eye".} \label{Fig.4}
\end{figure}

\end{document}